# Pressure-tuning topological phase transitions in Kagome superconductor $CsTi_3Bi_5$

*Wenfeng Wu[1,2]\*, Xiaocheng Bai[3]\*[#], Xianlong Wang[1,2], Dayong Liu[4]§, Zhi Zeng,[1,2] † and Liangjian Zou[1,2] ‡*

[1] Key Laboratory of Materials Physics, Institute of Solid State Physics, HFIPS, Chinese Academy of Sciences, Hefei 230031, China

[2] Science Island Branch of Graduate School, University of Science and Technology of China, Hefei 230026, China

[3] School of Science, Xi'an University of Posts and Telecommunications, Xi'an 710121, China

[4] Department of Physics, Nantong University, Nantong 230026, Jiangsu, China

Abstract: Recently, the Kagome metal $CsTi_3Bi_5$ has exhibited several novel quantum properties similar to $CsV_3Sb_5$, such as nontrivial topology, double-dome superconductivity, and flat band features. However, $CsTi_3Bi_5$ lacks the charge-density wave (CDW) present in $CsV_3Sb_5$, making the study of its emergence of double-dome superconductivity a focus of research. In this work, we have identified an order parameter, the three-band $Z_2$ topological index, that can describe the superconducting phase diagram of $CsTi_3Bi_5$ under pressure. Its evolution with pressure follows the expected behavior for superconductivity. Furthermore, the results of the Fermi surface under pressure reveal the potential presence of a Lifshitz transition in the vicinity of the vanishing point of the superconducting temperature change with pressure in $CsTi_3Bi_5$. These results indicate that the superconducting behavior of $CsTi_3Bi_5$ under pressure is caused by changes in the electronic structure leading to alterations in the topological properties, provide new insights and approaches for understanding the superconducting phenomenon in Kagome metals.

**Corresponding Author**

[#] xcbai@mail.ustc.edu.cn

§ dyliu@ntu.edu.cn

† zou@theory.issp.ac.cn

‡ zzeng@theory.issp.ac.cn

Currently topological superconductor is the most focusing candidate for quantum computation. In the past years it has anticipated that dope- or pressure-tuning topological insulators could become topological superconductor [1–5]. However, the carrier concentration or structure variation may modulate the topology of the systems, whether it remains topological in the superconductive phase still under debated, since the doping or pressure usually tunes the system out of the topological range. Recent discovery of superconductivity in Kagome metals $AV_3Sb_5$ (A = Cs, Rb) has stirred great interest on the interplays among topology, superconductivity, chiral charge density wave or charge order and structural distortion, and giant anomalous Hall effect [6–15]. Especially a lot of attention has been paid on the coexistence of superconductivity and charge density wave [16–19] and associated structural distortion, obscuring the effect of interplay between the topology and superconductivity.

Meanwhile, another kind Kagome metal $CsTi_3Bi_5$ was also found to be superconducting [20–22]. More interestingly, $CsTi_3Bi_5$ in the ambient pressure is a $Z_2$ topological metal [21–25]. In comparison with $AV_3Sb_5$, $CsTi_3Bi_5$ does not exhibit a charge density wave (CDW) phase under ambient pressure, which allows for investigating its interplay between topological and superconducting properties without the interference of structural instability [23,26]. It also demonstrates the absence of the structural instability up to the high pressure of 40 GPa [27]. This brings $CsTi_3Bi_5$ great superior to $AV_3Sb_5$ to study the possible topological superconductor. In comparison with $AV_3Sb_5$, the theoretical and experimental efforts paid for understanding the properties of $CsTi_3Bi_5$ are far from well done, leaving many open questions, such as, under ambient pressure or small hydrostatic pressure, whether $CsTi_3Bi_5$ is superconducting or not [21,23,28]; whether its superconducting phase is topological superconducting state, and how the electronic states and topology evolve with increasing applied hydrostatic pressure, and so on.

In this letter, we present the systematical study of the evolutions of the structural stability, electronic states and topology of $CsTi_3Bi_5$ with the increasing hydrostatic pressure up to 40 GPa. We find that the crystal structure $CsTi_3Bi_5$ with the space group P6/mmm is stable at the pressure smaller than 40 GPa. When the pressure increases from ambient to 40 GPa, $CsTi_3Bi_5$ undergoes a series $Z_2$ topological states in superconducting phases and trivial states in the normal phases. The symmetry-protected surface states also appear in the $Z_2$ topological phases. The Fermi surfaces also experience considerably changes with increasing pressure. These results clearly suggest that $CsTi_3Bi_5$ is possibly a realistic topologically-nontrivial superconductor, and hydrostatic pressure-could tune the topological states crossovers.

*Pressure-dependent atomic structures*: As shown in FIG. 1(a), the $CsTi_3Bi_5$ is a Kagome lattice material with the space group P6/mmm (No. 191). Specifically, Ti atoms reside in a two-dimensional plane, with one type of Bi atoms located in the same plane, while the other type of Bi atoms are perpendicular to the plane. Cs atoms serve as the cage-like framework of the material. The behavior of $CsTi_3Bi_5$ under pressures of 0 - 40 GPa, as mentioned earlier, does not exhibit any structural instability. Therefore, we used experimental lattice parameters to calculate its crystal structure under pressure [29], and the results are shown in FIG. 1(c). It is worth noting that, in the low-pressure range, the interlayer lattice constant $c$ is more sensitive to pressure compared to the in-plane lattice constant $a$. In this stage, the volume shows the most drastic change with pressure, which is characteristic of layered materials. At higher pressures, the rate of change of both $a$ and $c$ with pressure becomes nearly constant.

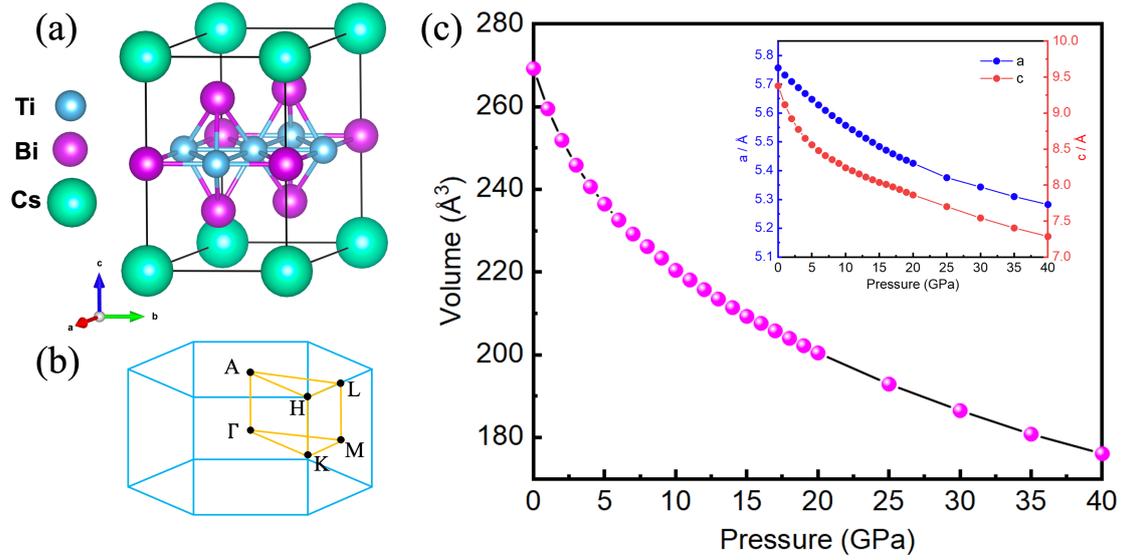

FIG. 1 (a) Crystal structure of $CsTi_3Bi_5$ at 0 GPa. (b) First Brillouin zone of $CsTi_3Bi_5$ with high-symmetry points labeled. (c) Variation of volume and lattice constants of $CsTi_3Bi_5$ with pressure.

*Topological Phase Diagram:* Under pressure of 0 – 40 GPa, there exist continuous band gaps between $\varepsilon$ and $\delta$ bands, as well as between the $\delta$ and $\gamma$ bands in $CsTi_3Bi_5$ and this will be discussed further in the electronic structure section. Furthermore, $CsTi_3Bi_5$ possesses time-reversal and inversion symmetries. Thus, we use the $Z_2$ topological invariants of the $\varepsilon$, $\delta$, and $\gamma$ bands to characterize the topological properties of $CsTi_3Bi_5$, as shown in FIG. 2, and the specific positions of these three bands can be found in FIG. 4 (a). At 0 GPa, we obtain $Z_2$ topological invariants for the three bands as (1 1 1), see FIG. 3 (a), consistent with previous reports [22,23]. It should be noted that due to the controversy surrounding the superconductivity of $CsTi_3Bi_5$ near 0 GPa, we choose the pressure range of 1 - 40 GPa to discuss its relationship with topology. We find that within this range, at least one of the $Z_2$ topological invariants for these three bands is 1, indicating that the system is a $Z_2$ topological metal from 1 to 40 GPa. In addition, at pressures from 1 to 11.5 GPa, the $Z_2$ topological invariant corresponding to the $\gamma$ band is 0, i.e., (1 1 0), which we refer to as topological metal II (TM II). When the pressure reaches 12 GPa, the system enters the (0 1 0) phase (TM III), and at 13 GPa, it enters the (0 1 1) phase (TM IV). The system remains in the TM III phase until the pressure increases to 34 GPa when it reenters the TM III phase and remains in this phase until 40 GPa. Our two-domed like topological phase diagram shows that the system undergoes topological phase transitions at 12 GPa and 35 GPa, which are consistent with the experimentally reported superconducting phase diagram. Moreover, the TM III phase occurring at 12 - 13 GPa and 34 - 40 GPa corresponds precisely to the pressure range in which the system is in a non-superconducting phase. This indicates that the topological phase transition induced by pressure may be a factor in the occurrence of superconductivity.

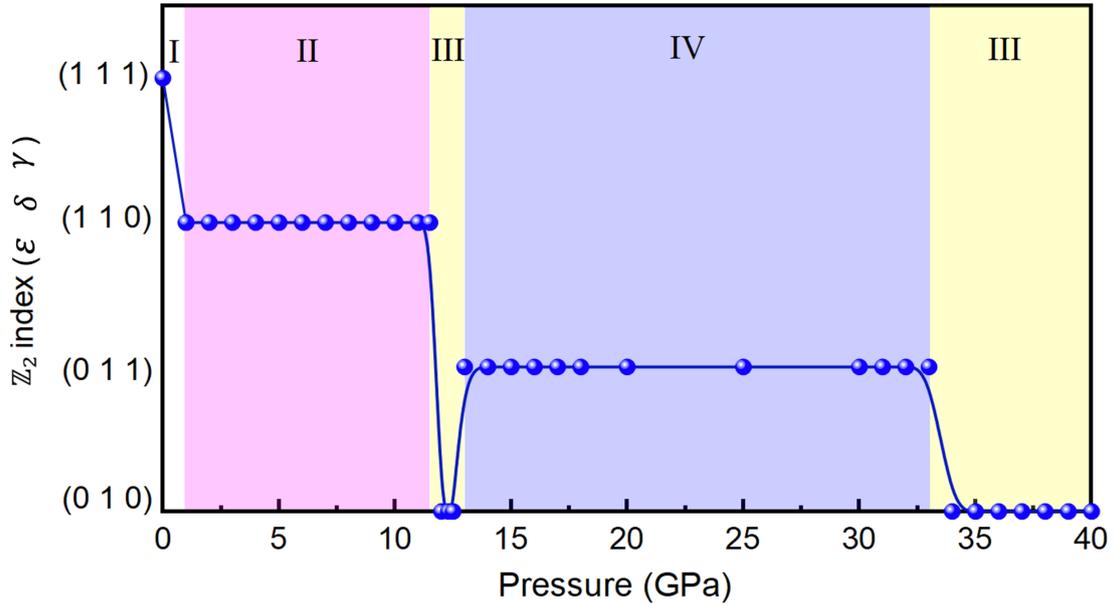

FIG. 2 The topological phase diagram of the $Z_2$ index of the $\varepsilon$, $\delta$, and $\gamma$ bands as a function of pressure. The pressure range in which the first phase exists is under debate in the superconducting phase diagram and it is separately labeled as TM I phase. The pink, yellow, and blue colors represent the TM II, TM III, and TM IV phases, respectively. Within the range of 1 - 40 GPa, the topological phase diagram can be roughly understood as having a two-dome structure.

By analyzing the product of parity at representative pressure points (5 GPa, 12.3 GPa, 30 GPa,

and 40 GPa) from 1 to 40 GPa, as shown in FIG. 3 (b) - (e). We find that as $CsTi_3Bi_5$ transitions from TM II to TM III, the parity of the $\varepsilon$ band at the Γ point changes sign, leading to a change in the product of parity and a transition of the $Z_2$ topological invariant from 1 to 0. Similarly, as the system transitions from TM III to TM IV, the parity of the $\gamma$ band at the L point changes sign, resulting in a transition of the $Z_2$ topological invariant from 0 to 1. When the pressure reaches around 34 GPa, the system transitions back from TM IV to TM III, and the parity of the $\gamma$ band at the L point changes sign again, causing the $Z_2$ topological invariant to transition from 1 back to 0. We find that the TM II-TM III topological phase transition is fundamentally different from the two subsequent transitions. The two topological phase transitions experienced by the system from 12 GPa onward are both related to the changes in the product of parity at the L point. This is due to its unique electronic structure.

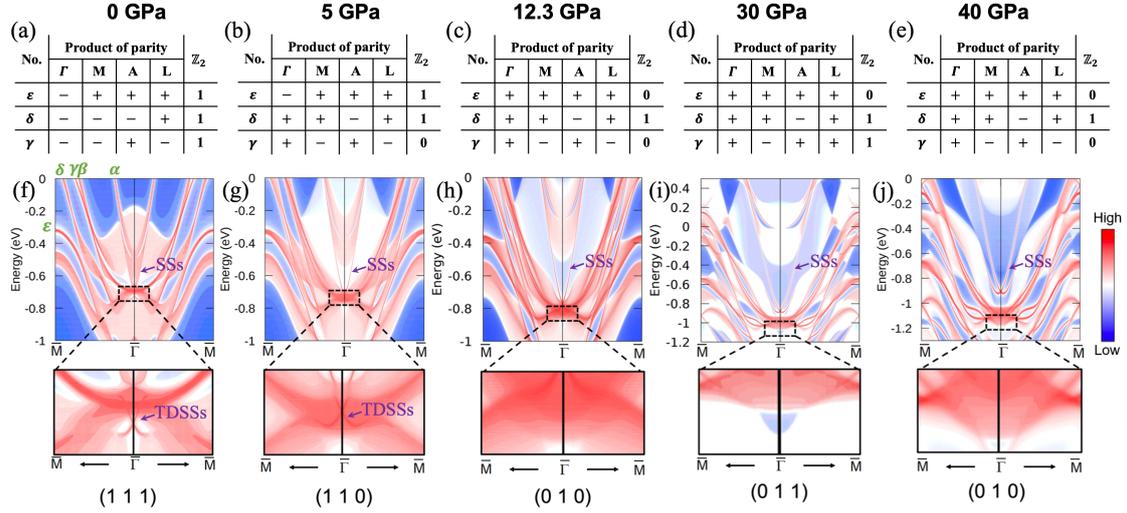

FIG. 3 The topological properties of the $CsTi_3Bi_5$ under pressures of 0, 5, 12.3, 30, and 40 GPa. (a)-(e) The parity of the $\varepsilon$, $\delta$, and $\gamma$ bands of $CsTi_3Bi_5$ at each pressure. (f) - (j) The surface states (SSs) and topologically nontrivial Dirac surface states (TDSSs) of five bands in $CsTi_3Bi_5$ at each pressure.

*Surface States:* After determining the topological classification of $CsTi_3Bi_5$, we have conducted a detailed study on the evolution of electronic states in surface of $CsTi_3Bi_5$ in different topological phases, see FIG. 3. By calculating the Green's function projection of the (001) surface spectra, at 0 GPa, the system exhibits topologically nontrivial Dirac surface states (TDSSs) between the $\varepsilon$ and $\delta$ bands, below the Fermi level by approximately 0.7 eV. Our calculations also identify surface states (SSs) near -0.6 eV; however, these states may overlap with the pure bulk states. These results are consistent with previous angle-resolved photoemission spectroscopy (ARPES) experiments and theoretical studies [21–23]. Under pressure, we found that the TDSSs move upward when the pressure increases to 5 GPa and completely disappear at 12.3 GPa until 40 GPa. This phenomenon corresponds to a change in the $Z_2$ topological invariant of the $\varepsilon$ band from 1 to 0.

*Electronic Structures:* The variation of the electronic structure of $CsTi_3Bi_5$ as a function of pressure will be discussed. At 0 GPa, considering the spin-orbit coupling (SOC), the band structure of the system is shown in FIG. 4(a). We find that at the M point, there are Van Hove singularities (vHS1 and vHS2) appearing around 0.2 eV and 0.7 eV above the Fermi level, respectively, and there are also some band inversion points along high-symmetry paths. The density of states (DOS) reveals that Ti elements predominantly occupy the vicinity of the Fermi surface, while Cs makes negligible contributions, we also can see a high peak around 0.5 eV below the Fermi level, which is contributed by Ti atoms and has localized electrons. Subsequently, we analyze the electronic structure at different pressure points (5, 12.3, 30, and 40 GPa) in the range of 1 - 40 GPa, corresponding to different topological phases. The results show that some Dirac points in the band structure without considering SOC (see FIG. S1) are opened when SOC is considered. In addition, for the results considering SOC, there is a continuous band gap between the $\delta$ and $\gamma$ bands. Under pressure, the vHS1, which is farther away from the Fermi level, changes slightly, but for the vHS2, which originally exists in the $\gamma$ band at 0 GPa, it gradually moves to the $\delta$ band at 5 GPa and 12.3 GPa. At 30 GPa, the $\delta$ band at the M point is suppressed below the Fermi level, and the $\gamma$ band reappears with vHS characteristics. At 40 GPa, both bands are further suppressed, and at the M point, the $\gamma$ band goes below the Fermi level. In terms of the DOS under pressure, we observed a trend of downward energy shift in the peak corresponding to the flat band as the pressure increases.

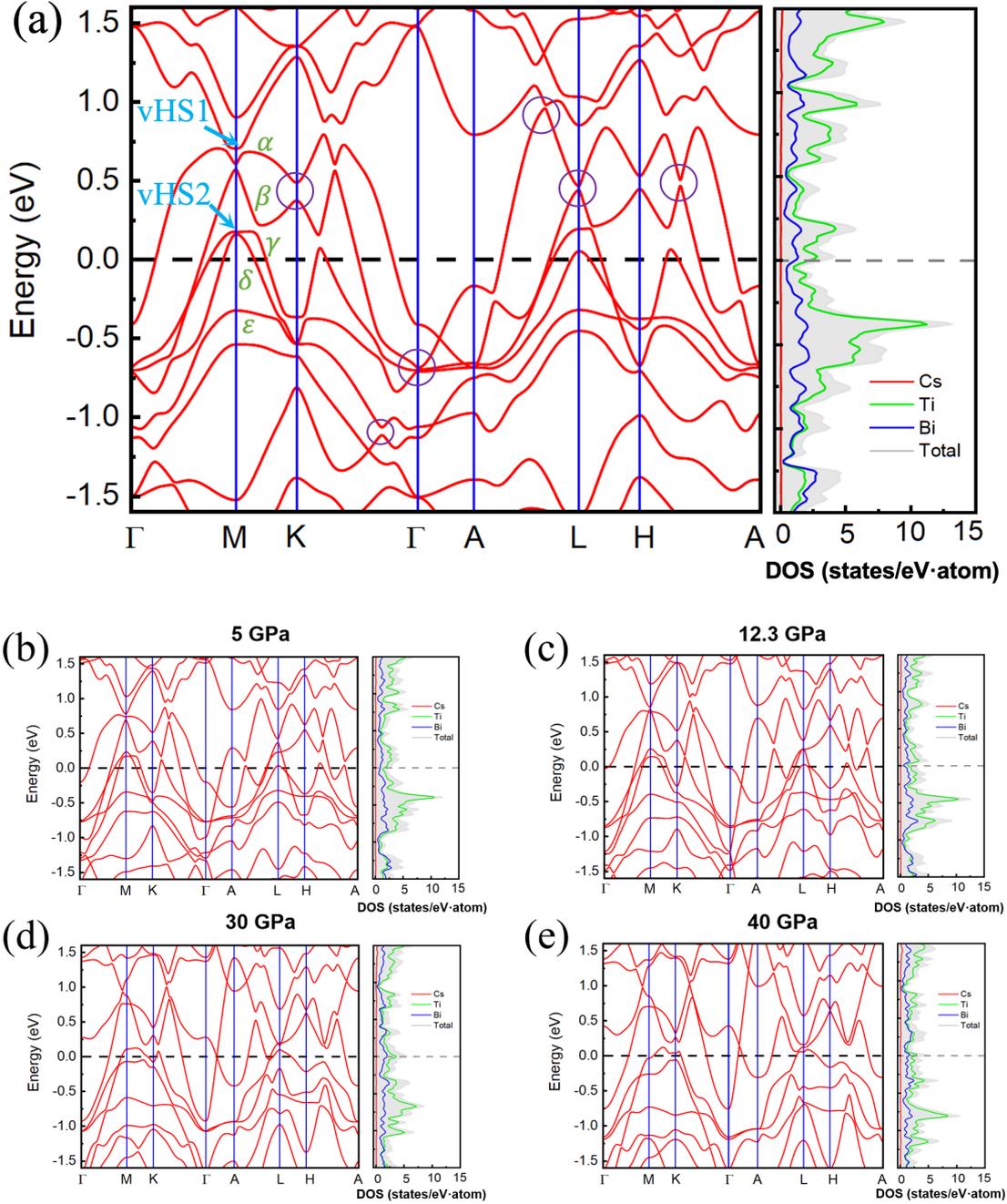

FIG. 4 (a) Band structure and DOS of CsTi$_3$Bi$_5$ considering SOC at 0 GPa. The blue markings indicate Van Hove singularities, while the specific positions of $\alpha$, $\beta$, $\varepsilon$, $\delta$, and $\gamma$ are marked in green. The purple circles represent energy bands exhibiting inverted characteristics. (b)-(e) The electronic structure of CsTi$_3$Bi$_5$ with SOC at typical pressures of 5, 12.3, 30, and 40 GPa. In terms of the DOS results, the red, green, blue, and gray lines represent the contributions from Cs, Ti, Bi atoms, and the total DOS, respectively.

The evolution of the Fermi surface at $k_z = 0$ under pressure was also investigated, as shown in FIG. 5. We found that as the pressure increases, the Fermi wave packet at the $\Gamma$ point gradually shrinks and disappears around 15 GPa. Around 30 GPa, the Fermi surface at the M point disappears due to the suppression of the $\delta$ band mentioned earlier to below the Fermi surface. At 40 GPa, further changes in the Fermi surface topology occur. These changes may lead to a Lifshitz phase transition similar to $CsV_3Sb_5$. From the evolution of the Fermi surface at $k_z = 0$, the pressure points where changes occur are around 15 GPa and 30 GPa, which have similarities with the pressure points of the topological phase transition described earlier. This suggests that the possible Lifshitz phase transition occurs simultaneously with the topological phase transition and the superconducting phase transition.

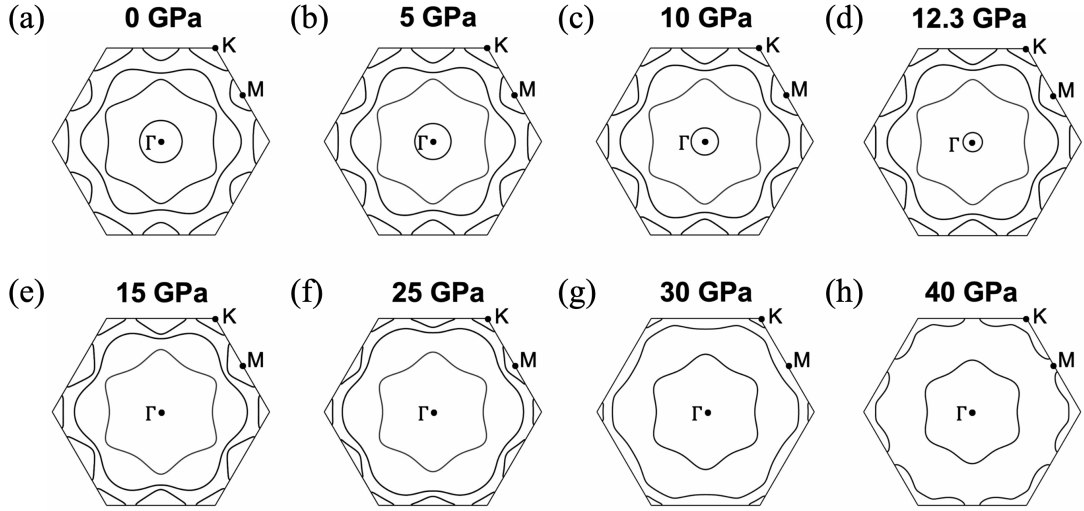

FIG. 5 Fermi surfaces of $CsTi_3Bi_5$ at pressures of 0, 5, 10, 12.3, 15, 25, 30, and 40 GPa at $k_z = 0$. The number of Fermi wave packets changes with increasing pressure in the ranges 12.5 - 15 GPa, 25 - 30 GPa, and 30 - 40 GPa.

*Discussions:* Here, we present a brief discussion on the relationship between topology and superconductivity in $CsTi_3Bi_5$. Our results indicate a correspondence between the topological phase diagram at pressures of 1 - 40 GPa and the two-domed superconducting phase diagram, suggesting the coexistence of topological and superconducting phases under pressure. Although the controversial existence of superconductivity in $CsTi_3Bi_5$ at 0 GPa [20,23], we infer from the aforementioned results that a superconducting phase should also be present at 0 GPa, as suggested by previous literature reporting a $T_c$ of 4.8 K. However, further research is needed to determine whether $CsTi_3Bi_5$ is a topological superconductor and to investigate potential pairing mechanisms. Finally, we discuss the potential existence of a Lifshitz phase transition. We note that the sister system $RbTi_3Bi_5$, after K atoms doping, exhibits a similar phenomenon where the top of the $\delta$ band is tuned from above to below $E_F$, resembling the suppression of the $\delta$ band below the Fermi level in $CsTi_3Bi_5$ under pressure of around 30 GPa. We speculate that K atom doping in $RbTi_3Bi_5$ induces lattice contraction, mimicking the effect of increased pressure. In fact, pressure manipulation is experimentally simpler, and we believe that pressure modulation of the Lifshitz phase transition holds promising prospects.

*Summary*: As a summary, we investigate the topological and electronic properties in $CsTi_3Bi_5$ under pressure using first-principles calculations. Our results reveal that the $Z_2$ topological invariant of the three bands near the Fermi surface exhibits a similar trend to the existing superconducting phase diagram with pressure. Additionally, the approximate location of the possible Lifshitz phase transition point coincides with the superconducting phase diagram. These findings highlight the importance of considering the topological and electronic properties in $CsTi_3Bi_5$ and call for further research in this area. This work is expected to provide valuable insights for understanding the emergence of a two-domed superconducting phase diagram in $CsTi_3Bi_5$.